\documentclass[pra,onecolumn,preprint]{revtex4}
\usepackage{graphicx}
\usepackage{amsfonts}
\usepackage{graphics}

\begin{document}

\preprint{}
\title[Efficient Simulation of Quantum States]{Efficient Simulation of
Quantum States Based on Classical Fields Modulated with Pseudorandom Phase
Sequences}
\author{Jian Fu, Shuo Sun}
\affiliation{State Key Lab of Modern Optical Instrumentation, Department of Optical
Engineering, Zhejiang University, Hangzhou 310027, China}

\begin{abstract}
We demonstrate that a tensor product structure could be obtained by
introducing pseudorandom phase sequences into classical fields with two
orthogonal modes. Using classical fields modulated with pseudorandom phase
sequences, we discuss efficient simulation of several typical quantum
states, including product state, Bell states, GHZ state, and W state. By
performing quadrature demodulation scheme, we could obtain the mode status
matrix of the simulating classical fields, based on which we propose a
sequence permutation mechanism to reconstruct the simulated quantum states%
\textbf{.} The research on classical simulation of quantum states is
important, for it not only enables potential practical applications in
quantum computation, but also provides useful insights into fundamental
concepts of quantum mechanics.
\end{abstract}

\date{today}
\startpage{1}
\email{jianfu@zju.edu.cn}
\maketitle

\section{Introduction \label{sec1}}

The classical simulation of quantum systems, especially of quantum
entanglement has been under investigation for a long time\ \cite%
{Cerf,Massar,Spreeuw}. In addition to potential practical applications in
quantum computation, research on classical simulation systems could help
understand some fundamental concepts in quantum mechanics. However, it has
been pointed out by several authors that classical simulation of quantum
systems exhibit exponentially scaling of physical resources with the number
of quantum particles \cite{Jozsa,Spreeuw}. In Ref. \cite{Spreeuw}, an
optical analogy of quantum systems is introduced, in which the number of
light beams and optical components required grows exponentially with the
number of cebits. In Ref. \cite{Vidal}, a classical protocol to efficiently
simulate any pure-state quantum computation is presented, yet the amount of
entanglement involved is restricted. In Ref. \cite{Jozsa}, it is elucidated
that in classical theory, the state space of a composite system is the
Cartesian product of subsystems, whereas in quantum theory it is the tensor
product. This essential distinction between Cartesian and tensor products is
precisely the phenomenon of quantum entanglement, and viewed as the origin
of the limitation of classical simulation of quantum systems.

In wireless and optical communications, orthogonal pseudorandom sequences
have been widely applied to Code Division Multiple Access (CDMA)
communication technology as a way to distinguish different users \cite%
{Viterbi,Peterson}. A set of pseudorandom sequences is generated by using a
shift register guided by a Galois field GF($p$), that satisfies orthogonal,
closure and balance properties \cite{Peterson}. In Phase Shift Keying (PSK)
communication systems, pseudorandom sequences are used to modulate the phase
of the electromagnetic/optical wave, where a pseudorandom sequence is mapped
to a pseudorandom phase sequence (PPS) values in $\left\{ 0,2\pi /p,\cdots
,2\pi \left( p-1\right) /p\right\} $. Guaranteed by the orthogonal property
of the PPS, different electromagnetic/optical waves could transmit in one
communication channel simultaneously with no crosstalk, and could be easily
distinguished by implementing a quadrature demodulation measurement \cite%
{Viterbi}.

In this paper, by introducing the PPSs into classical fields, we explore an
efficient simulation of quantum states based on classical fields with two
orthogonal modes. We demonstrate that $n$\textbf{\ }classical fields
modulated with $n$\ different PPSs can constitute a $2^{n}$-dimensional
Hilbert space that contains tensor product structure, which is similar with
quantum systems. In Ref. \cite{Fu1}, the efficient classical simulation of
Bell states and GHZ state has been introduced and both correlation analysis
and von Neumann entropy have been applied to characterize the simulation. In
this paper, by performing quadrature demodulation scheme, we could obtain
the mode status matrix of the simulating classical fields, based on which we
propose a sequence permutation mechanism to reconstruct the simulated
quantum states.\textbf{\ }Besides, classical simulation of some other
typical quantum states is discussed, including product state and W state. We
generalize our simulation and discuss the efficiency of our simulation in
the final end.

The paper is organized as follows: In Section \ref{sec2}, we introduce some
preparing knowledge needed later in this paper. In Section \ref{sec3}, the
existence of the tensor product structure in our simulation is demonstrated
and the classical simulation of several typical quantum states is analyzed.
In Section \ref{sec4}, a generalization of our simulation is proposed and
the efficiency of the simulation is discussed. Finally, we summarize our
conclusions in Section \ref{sec5}.

\section{Preparing Knowledge \label{sec2}}

In this section, we introduce some notation and basic results needed later
in this paper. We first introduce pseudorandom sequences and their
properties. Then we discuss the similarities between classical field and
single-particle quantum states. Finally, we introduce the scheme of
modulation and demodulation on classical fields with PPSs.

\subsection{Pseudorandom sequences and their properties \label{sec2.1}}

As far as we know, orthogonal pseudorandom sequences have been widely
applied to CDMA communication technology as a way to distinguish different
users \cite{Viterbi,Peterson}. A set of pseudorandom sequences is generated
by using a shift register guided by a Galois Field GF($p$), that satisfies
orthogonal, closure and balance properties. The orthogonal property ensures
that sequences of the set are independent and distinguished each other with
an excellent correlation property. The closure property ensures that any
linear combination of the sequences remains in the same set. The balance
property ensures that the occurrence rate of all non-zero-element equals
with each other, and the the number of zero-elements is exactly one less
than the other elements.

One famous generator of pseudorandom sequences is Linear Feedback Shift
Register (LFSR), which could produce a maximal period sequence, called
m-sequence \cite{Peterson}. We consider an m-sequence of period $N-1$ ($%
N=p^{s}$) generated by a primitive polynomial of degree $s$ over GF($p$).
Since the correlation between different shifts of an m-sequence is almost
zero, they can be used as different codes with their excellent correlation
property. In this regard, the set of $N-1$ m-sequences of length $N-1$ could
be obtained by cyclic shifting of a single m-sequence.

In this paper, we employ pseudorandom phase sequences (PPSs) with 4-ary
phase shift modulation, which is a well-known modulation format in wireless
and optical communications, including Orthogonal Quadrature Phase Shift
Keying (O-QPSK) and Minimum Phase Shift Keying (MSK) \cite{QPSK}. We first
propose a scheme to generate a PPS set $\Xi =\left\{ \lambda ^{\left(
0\right) },\lambda ^{\left( 1\right) },\ldots ,\lambda ^{\left( N-1\right)
}\right\} $ over GF($4$) \cite{Park}. $\lambda ^{\left( 0\right) }$ is an
all-$0$ sequence and other sequences can be generated by using the method as
follows:

(1) given a primitive polynomial of degree $s$ over GF($4$), a base sequence
of a length $4^{s}-1$ is generated by using LFSR;

(2) other sequences are obtained by cyclic shifting of the base sequence;

(3) by adding a zero-element to the end of each sequence, the occurrence
rates of all elements in all sequences are equal with each other;

(4) mapping the elements of the sequences to $\left[ 0,2\pi \right] $: $0$
mapping $0$, $1$ mapping $\pi /2$, $2$ mapping $\pi $, and $3$ mapping $3\pi
/2$.

Further, we define a map $f:\lambda \rightarrow e^{i\lambda }$ on the set of
$\Xi $, and obtain a new sequence set $\Omega =\left\{ \varphi ^{\left(
j\right) }\left\vert \varphi ^{\left( j\right) }=e^{i\lambda ^{\left(
j\right) }},\right. j=0,\ldots ,N-1\right\} $. According to the properties
of m-sequence, we can obtain following properties of the set $\Omega $, (1)
the closure property: the product of any sequences remains in the same set;
(2) the balance property: in exception to $\varphi ^{\left( 0\right) }$, any
sequence of the set $\Omega $ satisfy
\begin{equation}
\sum\limits_{k=1}^{N}e^{i\theta }\varphi _{k}^{\left( j\right)
}=\sum\limits_{k=1}^{N}e^{i\left( \lambda _{k}^{\left( j\right) }+\theta
\right) }=0,\forall \theta \in
\mathbb{R}
;  \label{eq0}
\end{equation}%
(3) the orthogonal property: any two sequences satisfy the following
normalized correlation

\begin{eqnarray}
E\left( {\varphi ^{\left( i\right) },\varphi ^{\left( j\right) }}\right) &=&%
\frac{1}{N}\sum\limits_{k=1}^{N}{\varphi _{k}^{\left( i\right) }\varphi
_{k}^{\left( j\right) \ast }}  \label{eq1} \\
&=&\left\{
\begin{array}{cc}
1, & i=j \\
0, & i\neq j%
\end{array}%
\right. .  \nonumber
\end{eqnarray}

In fact, the map $f$\ corresponds to the modulation of PPSs of $\Omega $ on
classical fields. According to the properties above, the classical fields
modulated with different PPSs become independent and distinguishable.

\subsection{Similarities between classical field and single-particle quantum
states \label{sec2.2}}

We note the similarities between Maxwell equation and Schr\"{o}dinger
equation. In fact, some properties of quantum information are wave
properties, where the wave need not be a quantum wave \cite{Spreeuw}.
Analogous to quantum states, classical fields also obey a superposition
principle, and could be transformed to any superposition state by unitary
transformations. Those analogous properties made the simulation of quantum
states using\textbf{\ }polarization or transverse modes of classical fields
possible \cite{Fu,dragoman,Lee}\textbf{. }

We first consider two orthogonal modes (polarization or transverse), which
are denoted by $\left\vert 0\right) $\ and $\left\vert 1\right) $\
respectively, as the classical simulation of quantum bits (qubits) $%
\left\vert 0\right\rangle $\ and\ $\left\vert 1\right\rangle $\ \cite%
{Spreeuw,Fu}.
\begin{equation}
\left\vert 0\right) =\left(
\begin{array}{c}
1 \\
0%
\end{array}%
\right) ,\left\vert 1\right) =\left(
\begin{array}{c}
0 \\
1%
\end{array}%
\right) .  \label{eq2}
\end{equation}%
Thus, any quantum state $|\varphi \rangle =\alpha |0\rangle +\beta |1\rangle
$\ can be simulated by a corresponding classical mode superposition field,
as follows
\begin{equation}
\left\vert \psi \right) =\alpha \left\vert 0\right) +\beta \left\vert
1\right) ,\left\vert \alpha \right\vert ^{2}+\left\vert \beta \right\vert
^{2}=1,\left( \alpha ,\beta \in
\mathbb{C}
\right) .  \label{eq3}
\end{equation}%
Obviously, all the mode superposition fields could span a Hilbert space,
where we can perform unitary transformations to transform the mode state.
For example, the unitary transformation $U\left( \chi ,\theta \right) $ is
defined
\begin{equation}
U\left( \chi ,\theta \right) =e^{i\chi \left( \sigma _{x}\cos \theta +\sigma
_{y}\sin \theta \right) }=\left(
\begin{array}{cc}
\cos \chi & -ie^{i\theta }\sin \chi \\
ie^{-i\theta }\sin \chi & \cos \chi%
\end{array}%
\right) ,  \label{eq4}
\end{equation}%
where $\sigma _{x}$, $\sigma _{y}$ are Pauli matrices. The modes $\left\vert
0\right) $ and $\left\vert 1\right) $ can be transformed to mode
superposition by using $U\left( \chi ,\theta \right) $, respectively, as
follows%
\begin{eqnarray}
U\left( \chi ,\theta \right) \left\vert 0\right) &=&\cos \chi \left\vert
0\right) +ie^{i\theta }\sin \chi \left\vert 1\right) ,  \label{eq6} \\
U\left( \chi ,\theta \right) \left\vert 1\right) &=&\cos \chi \left\vert
1\right) -ie^{-i\theta }\sin \chi \left\vert 0\right) .  \nonumber
\end{eqnarray}

Now, we consider some devices with one input and two outputs, such as beam
or mode splitters, which split one input field $\left\vert \psi _{in}\right)
=\alpha \left\vert 0\right) +\beta \left\vert 1\right) $ into two output
fields $\left\vert \psi _{out}^{\left( a\right) }\right) $ and $\left\vert
\psi _{out}^{\left( b\right) }\right) $. For the case of beam splitters, the
output fields are $\left\vert \psi _{out}^{\left( a\right) }\right)
=C_{a}\left( \alpha \left\vert 0\right) +\beta e^{i\phi _{a}}\left\vert
1\right) \right) $ and $\left\vert \psi _{out}^{\left( b\right) }\right)
=C_{b}\left( \alpha \left\vert 0\right) +\beta e^{i\phi _{b}}\left\vert
1\right) \right) $ with an arbitrary power ratio $\left\vert
C_{a}\right\vert ^{2}:\left\vert C_{b}\right\vert ^{2}$ between the output
beams, where $\phi _{a,b}$ are the additional phases due to the splitter.
For the case of mode splitters, the output fields are $\left\vert \psi
_{out}^{\left( a\right) }\right) =\alpha e^{i\phi _{a}}\left\vert 0\right) $
and $\left\vert \psi _{out}^{\left( b\right) }\right) =\beta e^{i\phi
_{b}}\left\vert 1\right) $, where $\phi _{a,b}$ are also the additional
phases. Conversely, the devices can act as beam or mode combiners in which
beams or modes from two inputs are combined into one output.

\subsection{Modulation and demodulation on classical fields with
pseudorandom phase sequences \label{sec2.3}}

We first consider the modulation process on a classical field with a PPS.
Similar to O-QPSK system, chosen a PPS $\lambda ^{\left( i\right) }$ in the
set of $\Xi $, the phase of the field could be modulated by a phase
modulator (PM) that controlled by a pseudorandom number generator (PNG), the
scheme is shown in Fig. 1. If the input is a single-mode field, it could be
transformed to mode superposition by performing a unitary transformation
after the modulation.

\begin{figure}
\centerline{\includegraphics[width=10cm]{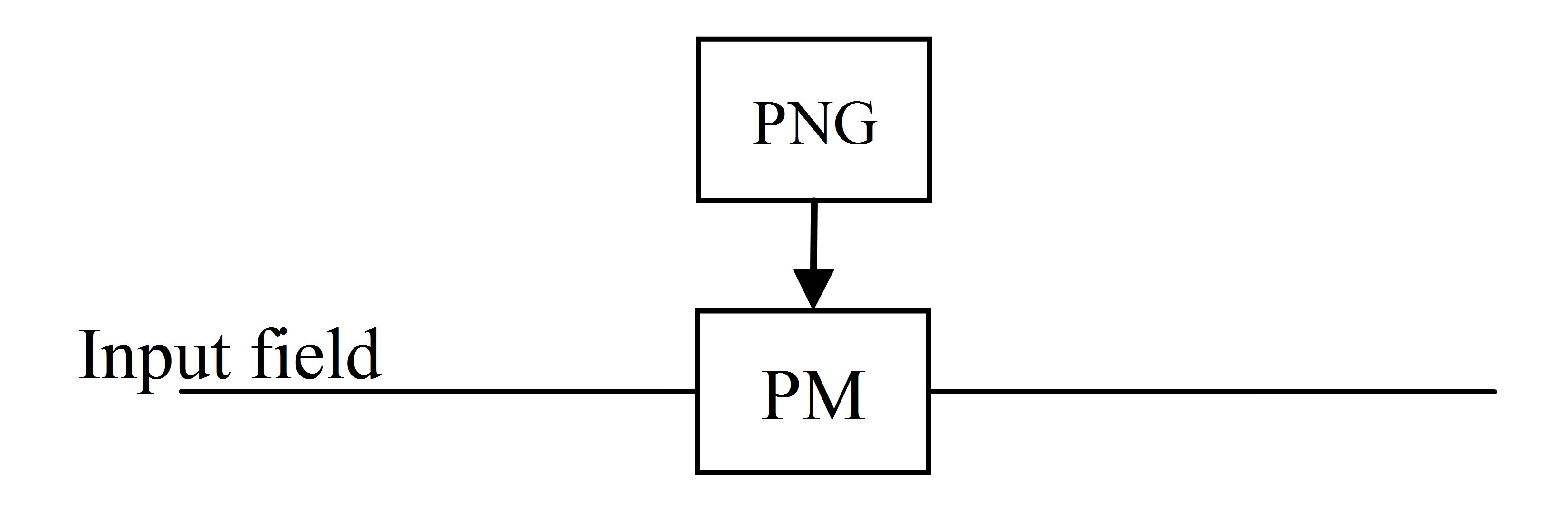}}
\caption{\label{fig1}The PPS encoding scheme for one
input field, where PNG denotes the pseudorandom number generator and PM
denotes the phase modulator.
}
\end{figure}

Then we consider the quadrature demodulation process of a modulated
classical field. Quadrature demodulation is a coherent detection process
that allows the simultaneous measurement of conjugate quadrature components
via homodyning the emerging beams with the input and reference fields by
using a balanced beam splitting, where the reference field is modulated with
a PPS $\lambda ^{\left( r\right) }$. The differenced signals of two output
detectors are then integrated and sampled to yield the decision variable. We
can express the demodulation process in mathematical form
\begin{eqnarray}
I\left( \lambda ^{\left( i\right) },\lambda ^{\left( r\right) }\right)  &=&%
\frac{1}{N}\sum_{k=1}^{N}\cos \left( \lambda _{k}^{\left( i\right) }-\lambda
_{k}^{\left( r\right) }\right)   \label{eq11} \\
&=&\left\{
\begin{array}{cc}
1, & i=r \\
0, & i\neq r%
\end{array}%
\right. ,  \nonumber
\end{eqnarray}%
where $\lambda _{k}^{\left( i\right) },\lambda _{k}^{\left( r\right) }$ are
the PPSs of the input and reference fields respectively. The output decision
variable is $1$\ if and only if $\lambda _{k}^{\left( i\right) },\lambda
_{k}^{\left( r\right) }$\ are equal; otherwise the output decision variable
is $0$. The results are guaranteed by the properties of PPSs. If the input
is a single-mode field, the scheme shown in Fig. 2 is employed to perform
quadrature demodulation. Otherwise the scheme shown in Fig. 3 is used, in
which the input field $\left\vert \psi _{i}\right) =e^{i\lambda ^{\left(
i\right) }}\left( \alpha _{i}\left\vert 0\right) +\beta _{i}\left\vert
1\right) \right) $ is first splitted into two fields $\alpha _{i}e^{i\lambda
^{\left( i\right) }}\left\vert 0\right) $ and $\beta _{i}e^{i\lambda
^{\left( i\right) }}\left\vert 1\right) $, and two coherent detection
processes are then performed on the two fields respectively. Noteworthily,
the modes of the reference fields must be consistent with the two output
fields. Thus there are two output decision variables $\tilde{\alpha}$\ and $%
\tilde{\beta}$, which correspond to the modes $\left\vert 0\right) $\ and $%
\left\vert 1\right) $, respectively. We define $\left( \tilde{\alpha},\tilde{%
\beta}\right) $ as the mode status. Besides the quadrature demodulation, we
can also easily measure the amplitudes of the fields $\left\vert \alpha
_{i}\right\vert ,\left\vert \beta _{i}\right\vert $ after mode spitting in
the scheme.

\begin{figure}
\centerline{\includegraphics[width=10cm]{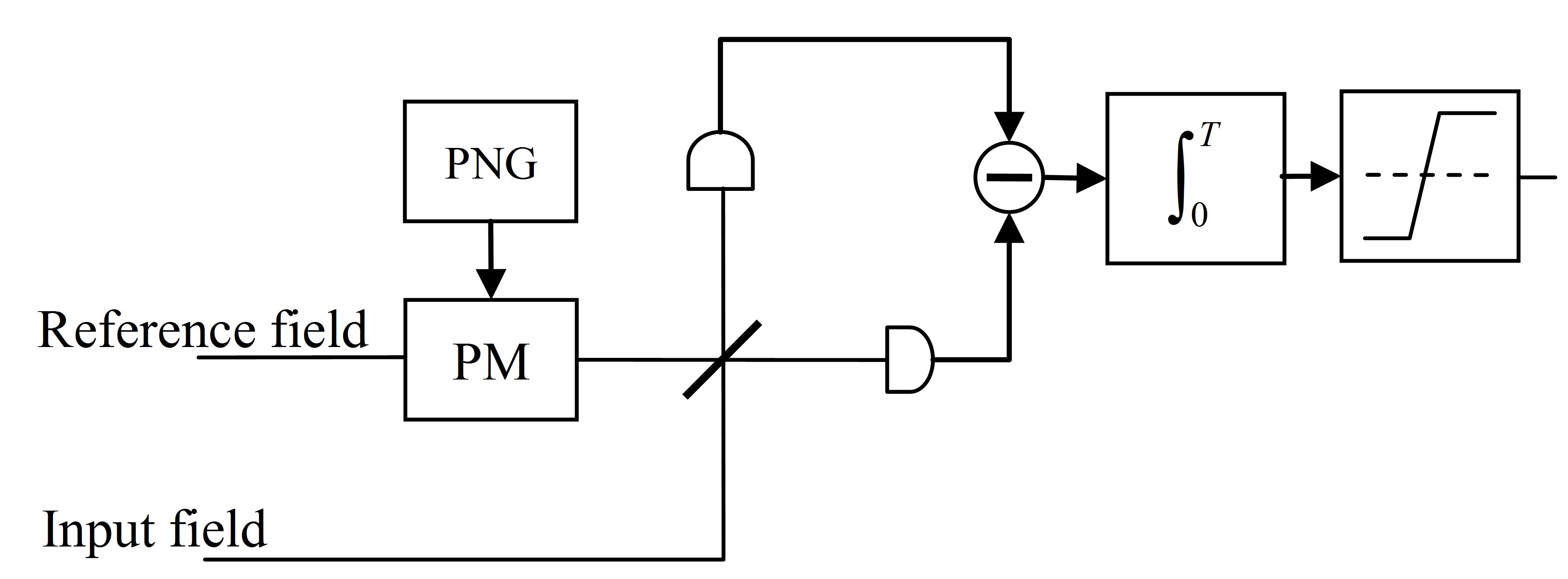}}
\caption{\label{fig2}The PPS quadrature demodulation
scheme for one input field.
}
\end{figure}

\begin{figure}
\centerline{\includegraphics[width=10cm]{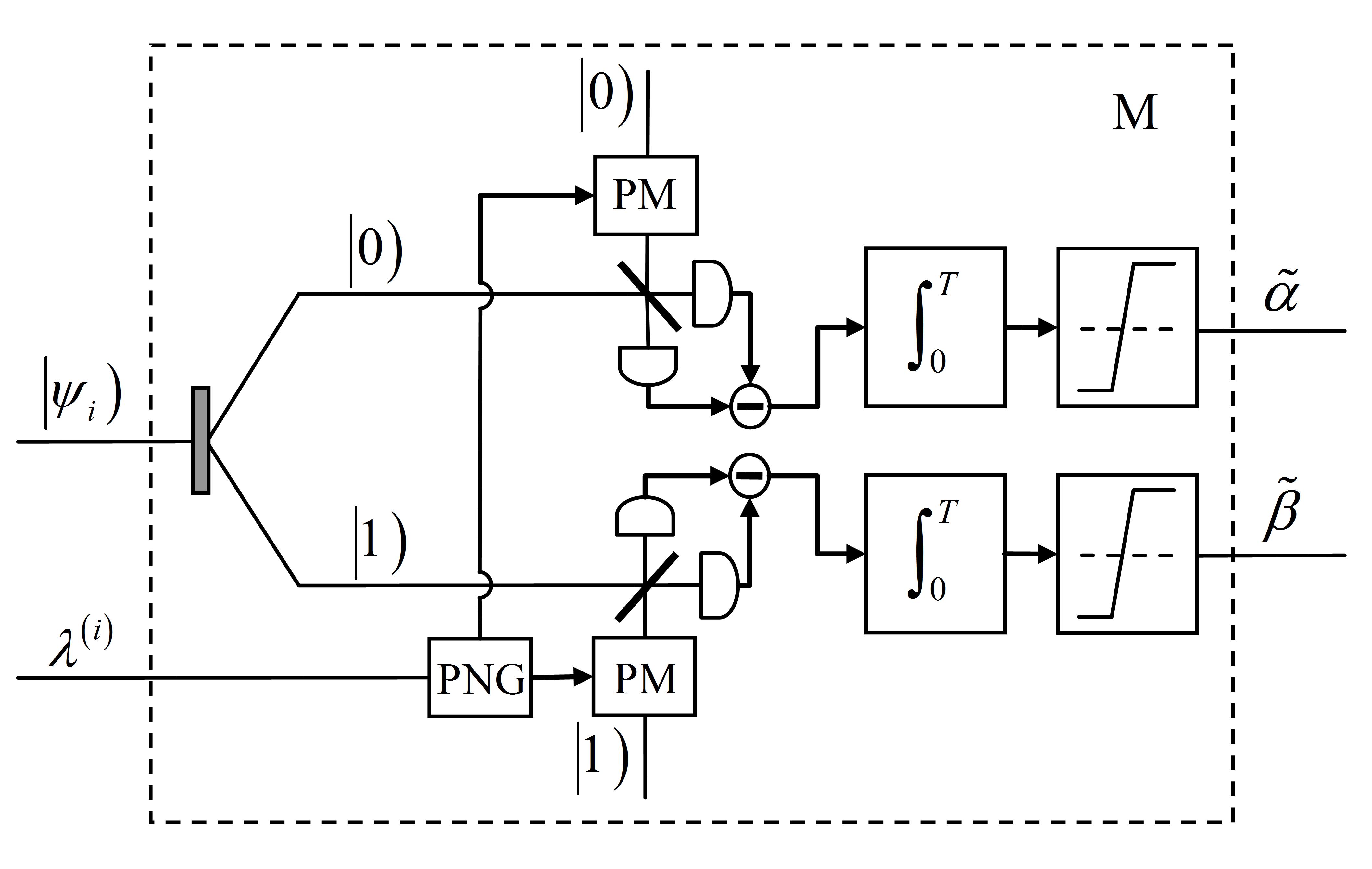}}
\caption{\label{fig3}The PPS quadrature demodulation
scheme for one field with two orthogonal modes, where the gray block denote
the mode splitter.
}
\end{figure}

\section{Classical simulation of multiparticle quantum states \label{sec3}}

We discuss simulation of multiparticle quantum states using classical fields
modulated with PPSs in this section. We first demonstrate that $n$ classical
fields modulated with $n$\ different PPSs could constitute a similar $2^{n}$%
-dimensional Hilbert space that contains a tensor product structure.
Besides, by performing quadrature demodulation scheme, we could obtain the
mode status matrix of the simulating classical fields, based on which we
propose a sequence permutation mechanism to reconstruct the simulated
quantum states\textbf{. }The classical simulation of some typical quantum
states is discussed, including product state, Bell states, GHZ state and W
state.

\subsection{Classical fields modulated with pseudorandom phase sequences and
their tensor product structure \label{sec3.1}}

For convenience, here we consider two classical fields modulated with PPSs
and their tensor product structure. Chosen two PPSs of $\lambda ^{\left(
a\right) }$ and $\lambda ^{\left( b\right) }$ from the set $\Xi $, any two
fields modulated with the PPSs can be expressed as follows,
\begin{eqnarray}
\left\vert \psi _{a}\right) &=&e^{i\lambda ^{\left( a\right) }}\left( \alpha
_{a}\left\vert 0\right) +\beta _{a}\left\vert 1\right) \right) ,
\label{eq13} \\
\left\vert \psi _{b}\right) &=&e^{i\lambda ^{\left( b\right) }}\left( \alpha
_{b}\left\vert 0\right) +\beta _{b}\left\vert 1\right) \right) .  \nonumber
\end{eqnarray}%
We define the inner product of two fields $\left\vert \psi _{a}\right) $ and
$\left\vert \psi _{b}\right) $ as follows,%
\begin{equation}
\left( \psi _{a}|\psi _{b}\right) =\frac{1}{N}\sum\limits_{k=1}^{N}e^{i%
\left( \lambda _{k}^{\left( a\right) }-\lambda _{k}^{\left( b\right)
}\right) }\left( \alpha _{b}^{\ast }\alpha _{a}+\beta _{b}^{\ast }\beta
_{a}\right) .  \label{eq14}
\end{equation}%
According to the properties of the PPSs, we can easily obtain%
\begin{equation}
\left( \psi _{a}|\psi _{b}\right) =\left\{
\begin{array}{cc}
1, & a=b \\
0, & a\neq b%
\end{array}%
\right. ,  \label{eq15}
\end{equation}%
which shows that two fields modulated with two different PPSs are
orthogonal. The orthogonal property supports to construct the tensor product
structure of multiple fields.

Assume two Hilbert spaces $w$ and $v$ spanned by the states $\left\vert \psi
_{a}\right) $ and $\left\vert \psi _{b}\right) $, any linear combinations of
the elements in the direct product space of $w\otimes v$ remain in the same
space. We define the two orthogonal modes modulated with the PPS $\lambda
^{\left( a\right) }$ as the orthonormal bases of the space of $w$, expressed
as $\left\vert 0_{a}\right) \equiv e^{i\lambda ^{\left( a\right)
}}\left\vert 0\right) $ and $\left\vert 1_{a}\right) \equiv e^{i\lambda
^{\left( a\right) }}\left\vert 1\right) $. Using the same notion, the
orthonormal bases of the space of $v$ are expressed as $\left\vert
0_{b}\right) \equiv e^{i\lambda ^{\left( b\right) }}\left\vert 0\right) $
and $\left\vert 1_{b}\right) \equiv e^{i\lambda ^{\left( b\right)
}}\left\vert 1\right) $.\ The four orthonormal bases are thus independent
and distinguishable. Then the orthonormal bases of the direct product space
of $w\otimes v$ can be expressed as $\left\{ \left\vert 0_{a}\right) \otimes
\left\vert 0_{b}\right) ,\left\vert 0_{a}\right) \otimes \left\vert
1_{b}\right) ,\left\vert 1_{a}\right) \otimes \left\vert 0_{b}\right)
,\left\vert 1_{a}\right) \otimes \left\vert 1_{b}\right) \right\} $.
Further, we can obtain the following tensor product properties,

(1) for any scalar $z$, the elements $\left\vert \psi _{a}\right)
,\left\vert \psi _{b}\right) $ in the spaces of $w$ and $v$, respectively,
satisfy
\begin{equation}
z\left( \left\vert \psi _{a}\right) \otimes \left\vert \psi _{b}\right)
\right) =\left( z\left\vert \psi _{a}\right) \right) \otimes \left\vert \psi
_{b}\right) =\left\vert \psi _{a}\right) \otimes \left( z\left\vert \psi
_{b}\right) \right) ;  \label{eq16}
\end{equation}

(2) in the space of $w\otimes v$, the direct product of the combinations of
elements is equal to the combination of the direct products of elements,
\begin{eqnarray}
\left( \left\vert \psi _{a}\right) +\left\vert \psi _{a}^{\prime }\right)
\right) \otimes \left( \left\vert \psi _{b}\right) +\left\vert \psi
_{b}^{\prime }\right) \right) &=&\left\vert \psi _{a}\right) \otimes
\left\vert \psi _{b}\right) +\left\vert \psi _{a}\right) \otimes \left\vert
\psi _{b}^{\prime }\right)  \label{eq17} \\
&&+\left\vert \psi _{a}^{\prime }\right) \otimes \left\vert \psi _{b}\right)
+\left\vert \psi _{a}^{\prime }\right) \otimes \left\vert \psi _{b}^{\prime
}\right) .  \nonumber
\end{eqnarray}%
Using the same notion, we can construct a $2^{n}$-dimensional direct product
space of $\left\vert \psi \right) ^{\otimes n}\equiv \left\vert \psi
_{1}\right) \otimes \ldots \otimes \left\vert \psi _{n}\right) $\ by using
the mode superposition of $n$\ classical fields $\left\vert \psi _{1}\right)
,\ldots ,\left\vert \psi _{n}\right) $\ modulated with $n$\ PPSs.

Quantum entanglement is only defined for Hilbert spaces that have a rigorous
tensor product structure in terms of subsystems. Here we construct a similar
structure of multiple classical fields, which is the basis of efficient
classical simulation of quantum entanglement.

\subsection{\textbf{R}econstruction of quantum states based on the
simulating classical fields \label{sec3.2}}

We have discussed quadrature demodulation process in Sec. \ref{sec2.3}. Here
we discuss how to reconstruct quantum state based on the simulating
classical fields with the help of quadrature demodulation.

First, we consider the general form of $n$ classical fields modulated with
PPSs $\left\{ \lambda ^{\left( 1\right) },\ldots ,\lambda ^{\left( n\right)
}\right\} $ chosen from the set $\Xi $,
\begin{eqnarray}
\left\vert \psi _{1}\right) &=&\frac{1}{C_{1}}\left[ \left( e^{i\lambda
^{\left( a\right) }}+...+e^{i\lambda ^{\left( b\right) }}\right) \left\vert
0\right) +\left( e^{i\lambda ^{\left( c\right) }}+...+e^{i\lambda ^{\left(
d\right) }}\right) \left\vert 1\right) \right] ,  \label{eq19} \\
&&......  \nonumber \\
\left\vert \psi _{n}\right) &=&\frac{1}{C_{n}}\left[ \left( e^{i\lambda
^{\left( e\right) }}+...+e^{i\lambda ^{\left( f\right) }}\right) \left\vert
0\right) +\left( e^{i\lambda ^{\left( g\right) }}+...+e^{i\lambda ^{\left(
h\right) }}\right) \left\vert 1\right) \right] ,  \nonumber
\end{eqnarray}%
where $C_{i}$ are the normalized coefficients and $a,\ldots ,h$ are the
sequence numbers. It is noteworthy that although multiple PPSs are
superimposed on both modes of the fields, all of the PPSs could be
demodulated and discriminated by performing the quadrature demodulation
introduced in Sec. \ref{sec2.3}, which has already been verified by many
actual communication systems \cite{QPSK}.

Now we propose a scheme, as shown in Fig. 4, to perform the quadrature
demodulation introduced in Sec. \ref{sec2.3}. In the scheme, quadrature
demodulations are performed on each field, in which the reference PPSs are
ergodic on $\left\{ \lambda ^{\left( 1\right) },\ldots ,\lambda ^{\left(
n\right) }\right\} $. Thus a mode status matrix $M\left( \tilde{\alpha}%
_{i}^{j},\tilde{\beta}_{i}^{j}\right) $, as shown in Fig. 5, could be
obtained by performing $n$ quadrature demodulations on the $n$ classical
fields. Of the matrix $M\left( \tilde{\alpha}_{i}^{j},\tilde{\beta}%
_{i}^{j}\right) $, each element is the mode status of the $i$th classical
field when the reference PPS is $\lambda ^{\left( j\right) }$. The element
takes one of four possible values: $\left( 1,0\right) ,\left( 0,1\right)
,\left( 1,1\right) $ or $0$, denote that the PPS $\lambda ^{\left( j\right)
} $ is modulated on mode $\left\vert 0\right) $ of the $i$th classical
field, on mode $\left\vert 1\right) $, on both $\left\vert 0\right) $ and $%
\left\vert 1\right) $, on neither $\left\vert 0\right) $ nor $\left\vert
1\right) $, respectively. It is noteworthy that different modulation of the $%
n$ classical fields correspond to different mode status matrixes, and vice
versa. Thus we obtain a one-to-one correspondence relationship between the $%
n $ classical fields and the mode status matrix. Besides, further discussion
will show that structure of quantum states and quantum entanglement could be
revealed in the mode status matrix, which means that a correspondence could
also be obtained between the mode status matrix and quantum states. Thus we
treat the mode status matrix as a bridge to connect the simulating fields
and the quantum states.

Now we focus on the correspondence between the mode status matrix and
quantum states, and consider how to reconstruct quantum states based on the
mode status matrix. We first transform the matrix $M\left( \tilde{\alpha}%
_{i}^{j},\tilde{\beta}_{i}^{j}\right) $ to a block diagonal matrix by
permuting the fields and the sequences, namely the rows and columns in the
matrix, respectively, and obtain a matrix expressed as similar to%
\begin{equation}
M\left( \tilde{\alpha}_{i}^{j},\tilde{\beta}_{i}^{j}\right) =\left(
\begin{array}{ccccccc}
M_{1}^{1} &  &  &  &  &  &  \\
& M_{2}^{2} & M_{2}^{3} &  &  &  &  \\
& M_{3}^{2} & M_{3}^{3} &  &  &  &  \\
&  &  & M_{4}^{4} & M_{4}^{5} & M_{4}^{6} &  \\
&  &  & M_{5}^{4} & M_{5}^{5} & M_{5}^{6} &  \\
&  &  & M_{6}^{4} & M_{6}^{5} & M_{6}^{6} &  \\
&  &  &  &  &  & \ddots%
\end{array}%
\right) .  \label{eq20}
\end{equation}

The structure of the matrix could clearly reveal the structure of the
simulated quantum state. Each diagonal block of the matrix denotes one
unreduced subsystem of the simulated state $\left\vert \Psi \right) $. Thus
the state $\left\vert \Psi \right) $ could be expressed as the direct
product of the unreduced states $\left\vert \Psi _{b}\right) $, as follows

\begin{equation}
\left\vert \Psi \right) =\prod_{b=1}^{m}\left\vert \Psi _{b}\right) ,
\label{eq21}
\end{equation}%
where $m$ denotes the number of the matrix blocks.

Now we consider how to reconstruct each $\left\vert \Psi _{b}\right) $ based
on each submatrix, respectively. As each submatrix is irreducible, each $%
\left\vert \Psi _{b}\right) $ corresponds to an entangled state. More
important, different entanglement structures correspond to different
structures of the unreduced submatrix. Thus a correspondence relationship
could be obtained between quantum entanglement and the unreduced mode status
matrix. In order to reconstruct the quantum entanglement state, an ergodic
ensemble of PPSs is required to obtain all possible base states. Thus we
propose a sequence permutation mechanism to reconstruct each $\left\vert
\Psi _{b}\right) $ based on each submatrix, which is one of the simplest
mechanisms for sequence ergodic ensemble. Assumed $\left\vert \Psi
_{b}\right) $ contains $l$ classical fields with $l$ PPSs, namely the
corresponding submatrix contains $l$ rows and $l$ columns, the sequence
permutation is arranged as
\begin{equation}
R_{1}=\left\{ \lambda ^{\left( 1\right) },\lambda ^{\left( 2\right) },\ldots
,\lambda ^{\left( l\right) }\right\} ,R_{2}=\left\{ \lambda ^{\left(
2\right) },\lambda ^{\left( 3\right) },\ldots ,\lambda ^{\left( 1\right)
}\right\} ,\ldots ,R_{l}=\left\{ \lambda ^{\left( l\right) },\lambda
^{\left( 1\right) },\ldots ,\lambda ^{\left( l-1\right) }\right\} .
\label{eq22}
\end{equation}%
In the mechanism, each $R_{r}$ corresponds to one selection from the
unreduced submatrix. We obtain a direct product of $l$ items for each $R_{r}$%
, and the simulated quantum state is the superposition of the $l$ product
items. Therefore we obtain%
\begin{equation}
\left\vert \Psi _{b}\right) =\frac{1}{C}\sum_{r=1}^{l}\prod%
\limits_{i=1}^{l}\left( \tilde{\alpha}_{i}^{j\left[ R_{r}^{i}\right]
}\left\vert 0\right) +\tilde{\beta}_{i}^{j\left[ R_{r}^{i}\right]
}\left\vert 1\right) \right) ,  \label{eq23}
\end{equation}%
where $C$ is the normalized coefficient, $\left( \tilde{\alpha}_{i}^{j\left[
R_{r}^{i}\right] },\tilde{\beta}_{i}^{j\left[ R_{r}^{i}\right] }\right) $ is
the mode status obtained from the submatrix $M\left( \tilde{\alpha}_{i}^{j},%
\tilde{\beta}_{i}^{j}\right) $, where $j\left[ R_{r}^{i}\right] $ denotes
the sequence number of the $i$th sequence in $R_{r}$.

It is noteworthy that the mechanism we proposed above is one of the feasible
ways to reconstruct the simulated quantum state based on the unreduced mode
status matrix. Other mechanisms might also work, as long as a sequence
ergodic ensemble is obtained in the mechanism. The sequence permutation
mechanism above could successfully reconstruct many quantum states,
including the product states, Bell states, GHZ states and W states. We will
discuss the related contents in next subsection.

\begin{figure}
\centerline{\includegraphics[width=10cm]{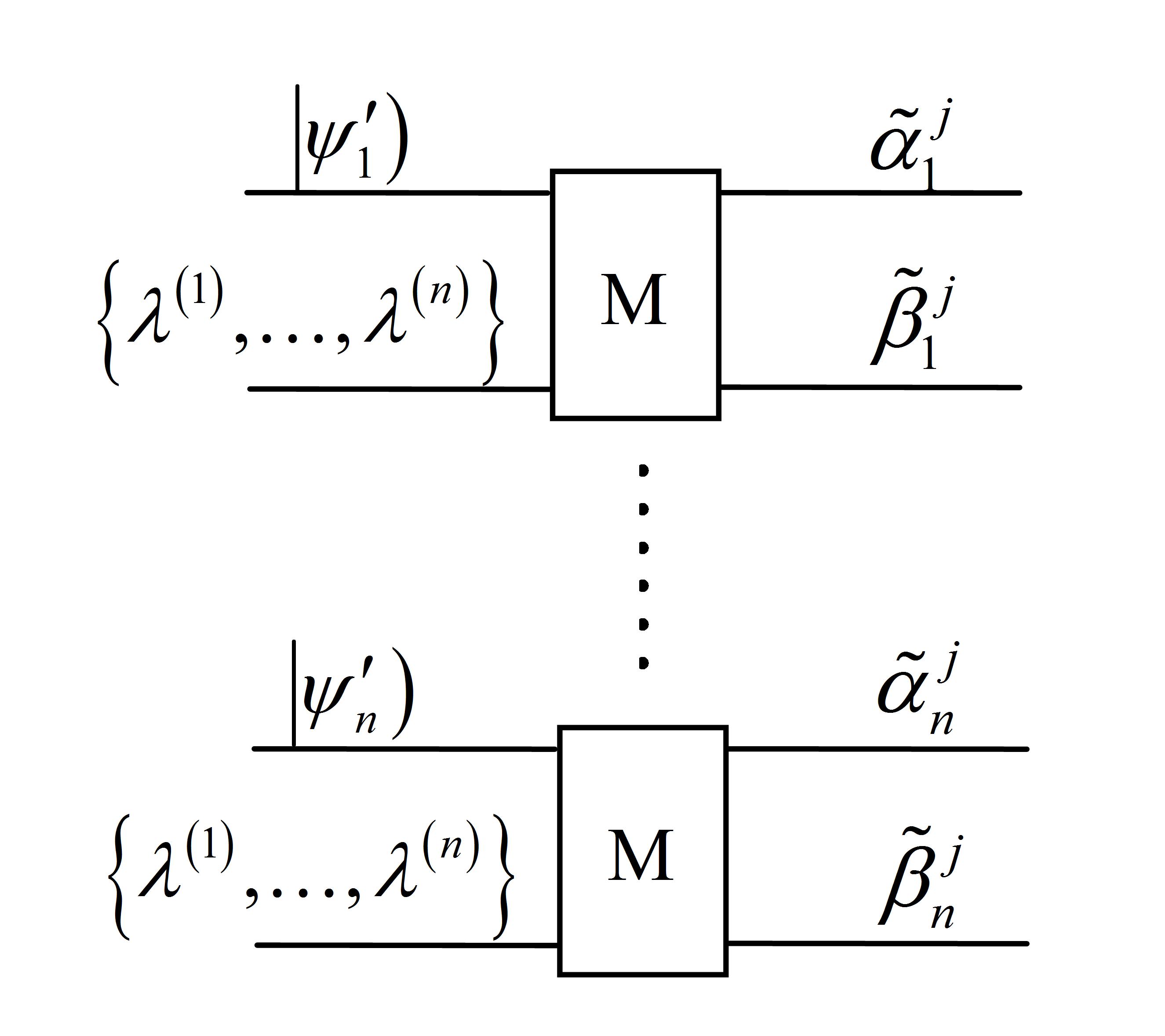}}
\caption{\label{fig4}The PPS quadrature demodulation
scheme for multiple input fields, where the M block is shown in Fig. 3.
}
\end{figure}

\begin{figure}
\centerline{\includegraphics[width=10cm]{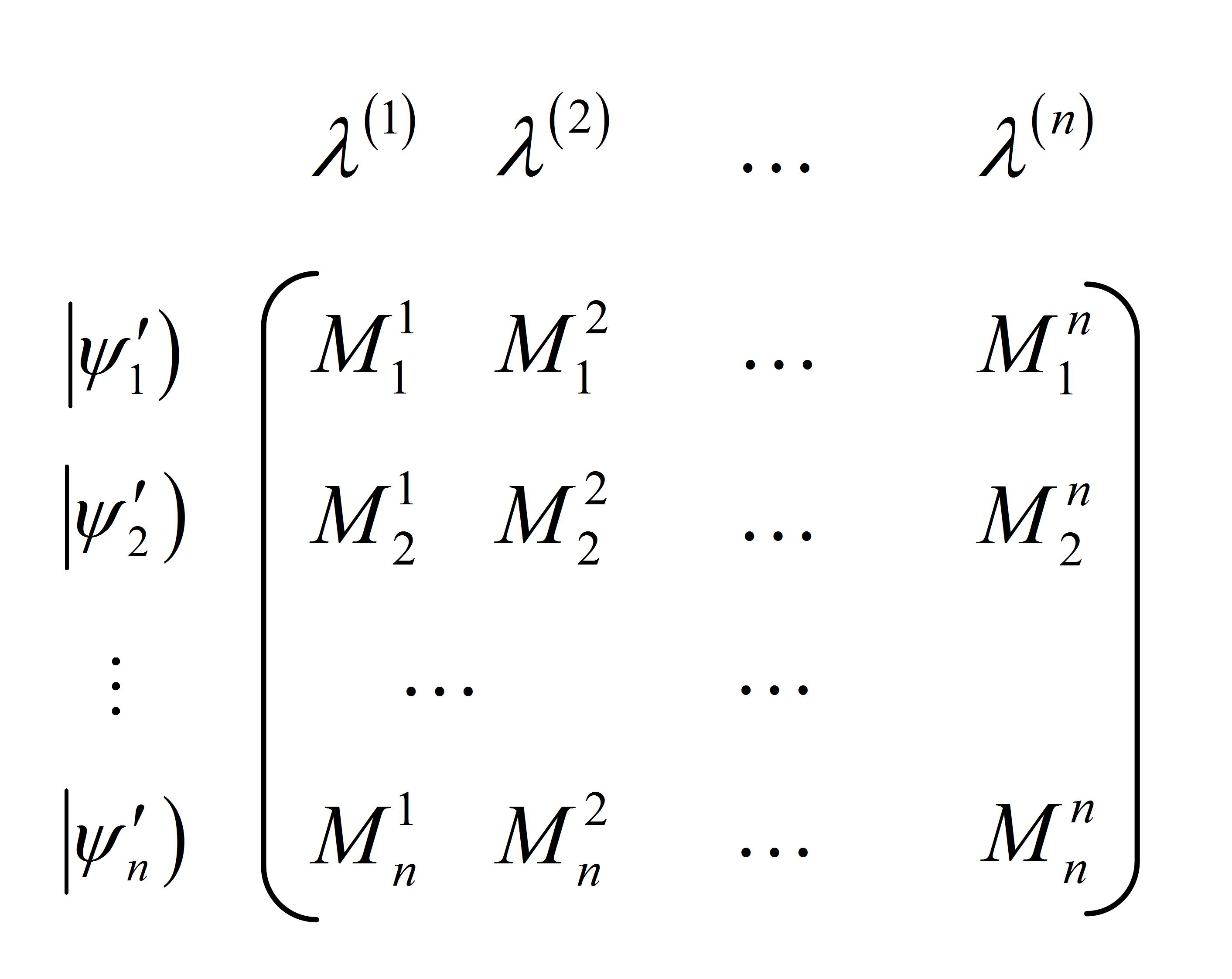}}
\caption{\label{fig5}The mode status matrix related
to the fields and PPSs, where $M_{i}^{j}$ is the element of $M\left( \tilde{%
\protect\alpha}_{i}^{j},\tilde{\protect\beta}_{i}^{j}\right) $ for the $i$th
classical field and the reference PPS $\protect\lambda ^{\left( j\right) }$.
}
\end{figure}

\subsection{Classical simulation of several typical quantum states \label%
{sec3.3}}

In this subsection, we discuss classical simulation of several typical
quantum states, including product state, Bell states, GHZ state and W state.

\subsubsection{Classical simulation of product state \label{sec3.3.1}}

First, we discuss classical simulation of $n$\ quantum product state. The
simulation fields are shown as follows
\begin{eqnarray}
\left\vert \psi _{1}\right) &=&\frac{e^{i\lambda ^{\left( 1\right) }}}{\sqrt{%
2}}\left( \left\vert 0\right) +\left\vert 1\right) \right) ,  \label{eq18} \\
&&......  \nonumber \\
\left\vert \psi _{n}\right) &=&\frac{e^{i\lambda ^{\left( n\right) }}}{\sqrt{%
2}}\left( \left\vert 0\right) +\left\vert 1\right) \right) .  \nonumber
\end{eqnarray}%
By employing the scheme as shown in Fig. 4, we obtain the mode status matrix
\begin{equation}
M\left( \tilde{\alpha}_{i}^{j},\tilde{\beta}_{i}^{j}\right) =\left(
\begin{array}{ccc}
\left( 1,1\right) &  & 0 \\
& \ddots &  \\
0 &  & \left( 1,1\right)%
\end{array}%
\right) ,  \label{eq24}
\end{equation}%
which demonstrates that each classical field is the superposition of two
orthogonal modes and no entanglement is involved. According to Eq. (\ref%
{eq21}), we obtain
\begin{eqnarray}
\left\vert \Psi \right) &=&\frac{1}{2^{n/2}}\left( \left\vert 0\right)
+\left\vert 1\right) \right) \otimes \ldots \otimes \left( \left\vert
0\right) +\left\vert 1\right) \right)  \label{eq25} \\
&=&\frac{1}{2^{n/2}}\left( \left\vert 0\ldots 0\right) +\left\vert 0\ldots
1\right) +\ldots +\left\vert 1\ldots 1\right) \right) ,  \nonumber
\end{eqnarray}%
where $\left\vert q_{1}\ldots q_{n}\right) \equiv \left\vert q_{1}\right)
\otimes \ldots \otimes \left\vert q_{n}\right) ,\left( q_{i}=0\ or\ 1\right)
$.

\subsubsection{Classical simulation of Bell states \label{sec3.3.2}}

Now we discuss classical simulation of one of the four Bell states $%
\left\vert \Psi ^{+}\right\rangle =\frac{1}{\sqrt{2}}\left( \left\vert
0_{a}\right\rangle \left\vert 0_{b}\right\rangle +\left\vert
1_{a}\right\rangle \left\vert 1_{b}\right\rangle \right) $, which contains
two classical fields as follows%
\begin{eqnarray}
\left\vert \psi _{a}\right) &=&\frac{1}{\sqrt{2}}\left( e^{i\lambda ^{\left(
a\right) }}\left\vert 0\right) +e^{i\lambda ^{\left( b\right) }}\left\vert
1\right) \right) ,  \label{eq26} \\
\left\vert \psi _{b}\right) &=&\frac{1}{\sqrt{2}}\left( e^{i\lambda ^{\left(
b\right) }}\left\vert 0\right) +e^{i\lambda ^{\left( a\right) }}\left\vert
1\right) \right) .  \nonumber
\end{eqnarray}%
By employing the scheme as shown in Fig. 4, we obtain the mode status matrix%
\begin{equation}
M\left( \tilde{\alpha}_{i}^{j},\tilde{\beta}_{i}^{j}\right) =\left(
\begin{array}{cc}
\left( 1,0\right) & \left( 0,1\right) \\
\left( 0,1\right) & \left( 1,0\right)%
\end{array}%
\right) .  \label{eq27}
\end{equation}%
We note that in this case, the mode status matrix is irreducible, which
corresponds to an entanglement state. According to the sequence permutation
mechanism, we obtain that $R_{1}=\{\lambda ^{\left( a\right) },\lambda
^{\left( b\right) }\}$ and $R_{2}=\{\lambda ^{\left( b\right) },\lambda
^{\left( a\right) }\}$. Based on the mode status matrix, for the selection
of $R_{1}$, we obtain $\left\vert 0\right) \otimes \left\vert 0\right) $;
for the selection of $R_{2}$, we obtain $\left\vert 1\right) \otimes
\left\vert 1\right) $. If we randomly choose one selection between $R_{1}$
and $R_{2}$, we could randomly obtain one result between $\left\vert
0\right) \otimes \left\vert 0\right) $ and $\left\vert 1\right) \otimes
\left\vert 1\right) $, which is similar with the case of quantum measurement
for the Bell state $\left\vert \Psi ^{+}\right\rangle $. We could
reconstruct the state based on the mode status matrix%
\begin{eqnarray}
\left\vert \Psi ^{+}\right) &=&\frac{1}{\sqrt{2}}\sum_{r=1}^{2}\left[ \left(
\tilde{\alpha}_{a}^{j\left[ R_{r}^{a}\right] }\left\vert 0\right) +\tilde{%
\beta}_{a}^{j\left[ R_{r}^{a}\right] }\left\vert 1\right) \right) \otimes
\left( \tilde{\alpha}_{b}^{j\left[ R_{r}^{b}\right] }\left\vert 0\right) +%
\tilde{\beta}_{b}^{j\left[ R_{r}^{b}\right] }\left\vert 1\right) \right) %
\right]  \label{eq28} \\
&=&\frac{1}{\sqrt{2}}\left( \left\vert 00\right) +\left\vert 11\right)
\right) .  \nonumber
\end{eqnarray}

In quantum mechanics, another Bell state $\left\vert \Phi ^{+}\right\rangle $
could be obtained from $\left\vert \Psi ^{+}\right\rangle $ by performing
the unitary transformation $\sigma _{x}:\left\vert 0\right\rangle
\leftrightarrow \left\vert 1\right\rangle $ on one of the particles. Using
the same method, we perform an unitary transformation on $\left\vert \psi
_{b}\right) $ to flip its modes $\left\vert 0\right) \leftrightarrow
\left\vert 1\right) $. Thus we obtain two classical fields as follows%
\begin{eqnarray}
\left\vert \psi _{a}\right) &=&\frac{1}{\sqrt{2}}\left( e^{i\lambda ^{\left(
a\right) }}\left\vert 0\right) +e^{i\lambda ^{\left( b\right) }}\left\vert
1\right) \right) ,  \label{eq29} \\
\left\vert \psi _{b}\right) &=&\frac{1}{\sqrt{2}}\left( e^{i\lambda ^{\left(
b\right) }}\left\vert 1\right) +e^{i\lambda ^{\left( a\right) }}\left\vert
0\right) \right) .  \nonumber
\end{eqnarray}%
By employing the scheme as shown in Fig. 4, we obtain the mode status matrix%
\begin{equation}
M\left( \tilde{\alpha}_{i}^{j},\tilde{\beta}_{i}^{j}\right) =\left(
\begin{array}{cc}
\left( 1,0\right) & \left( 0,1\right) \\
\left( 1,0\right) & \left( 0,1\right)%
\end{array}%
\right) .  \label{eq30}
\end{equation}%
According to the sequence permutation mechanism, here we obtain $%
R_{1}=\{\lambda ^{\left( a\right) },\lambda ^{\left( b\right) }\}$ and $%
R_{2}=\{\lambda ^{\left( b\right) },\lambda ^{\left( a\right) }\}$ again. As
the mode status matrix is different, for $R_{1}$, the result turns to be $%
\left\vert 0\right) \otimes \left\vert 1\right) $; for $R_{2}$, we obtain $%
\left\vert 1\right) \otimes \left\vert 0\right) $. If we randomly choose one
selection between $R_{1}$ and $R_{2}$, we could also randomly obtain one
result between $\left\vert 0\right) \otimes \left\vert 1\right) $ and $%
\left\vert 1\right) \otimes \left\vert 0\right) $, which is similar with the
case of quantum measurement for the Bell state $\left\vert \Phi
^{+}\right\rangle $. We could reconstruct the state
\begin{eqnarray}
\left\vert \Phi ^{+}\right) &=&\frac{1}{\sqrt{2}}\sum_{r=1}^{2}\left[ \left(
\tilde{\alpha}_{a}^{j\left[ R_{r}^{a}\right] }\left\vert 0\right) +\tilde{%
\beta}_{a}^{j\left[ R_{r}^{a}\right] }\left\vert 1\right) \right) \otimes
\left( \tilde{\alpha}_{b}^{j\left[ R_{r}^{b}\right] }\left\vert 0\right) +%
\tilde{\beta}_{b}^{j\left[ R_{r}^{b}\right] }\left\vert 1\right) \right) %
\right]  \label{eq31} \\
&=&\frac{1}{\sqrt{2}}\left( \left\vert 01\right) +\left\vert 10\right)
\right) .  \nonumber
\end{eqnarray}

For other two Bell states $\left\vert \Psi ^{-}\right) $ and $\left\vert
\Phi ^{-}\right) $, they could be obtained from $\left\vert \Psi ^{+}\right)
$ and $\left\vert \Phi ^{+}\right) $ by using a $\pi $ phase transformation.
However, they could not be distinguished from $\left\vert \Psi ^{+}\right) $
and $\left\vert \Phi ^{+}\right) $ unless an orthogonal projection
measurement is performed \cite{Fu1}.

\subsubsection{Classical simulation of GHZ state \label{sec3.3.3}}

For tripartite systems there are only two different classes of genuine
tripartite entanglement, the GHZ class and the W class \cite%
{Greenberger,Nielsen}. First we discuss the classical simulation of GHZ
state $\left\vert \Psi _{GHZ}\right\rangle =\frac{1}{\sqrt{2}}\left(
\left\vert 0_{a}\right\rangle \left\vert 0_{b}\right\rangle \left\vert
0_{c}\right\rangle +\left\vert 1_{a}\right\rangle \left\vert
1_{b}\right\rangle \left\vert 1_{c}\right\rangle \right) $, which contains
three classical fields as follows%
\begin{eqnarray}
\left\vert \psi _{a}\right) &=&\frac{1}{\sqrt{2}}\left( e^{i\lambda ^{\left(
a\right) }}\left\vert 0\right) +e^{i\lambda ^{\left( b\right) }}\left\vert
1\right) \right) ,  \label{eq32} \\
\left\vert \psi _{b}\right) &=&\frac{1}{\sqrt{2}}\left( e^{i\lambda ^{\left(
b\right) }}\left\vert 0\right) +e^{i\lambda ^{\left( c\right) }}\left\vert
1\right) \right) ,  \nonumber \\
\left\vert \psi _{c}\right) &=&\frac{1}{\sqrt{2}}\left( e^{i\lambda ^{\left(
c\right) }}\left\vert 0\right) +e^{i\lambda ^{\left( a\right) }}\left\vert
1\right) \right) .  \nonumber
\end{eqnarray}%
Performing the scheme as shown in Fig. 4, we obtain the mode status matrix%
\begin{equation}
M\left( \tilde{\alpha}_{i}^{j},\tilde{\beta}_{i}^{j}\right) =\left(
\begin{array}{ccc}
\left( 1,0\right) & \left( 0,1\right) & 0 \\
0 & \left( 1,0\right) & \left( 0,1\right) \\
\left( 0,1\right) & 0 & \left( 1,0\right)%
\end{array}%
\right) .  \label{eq33}
\end{equation}%
According to the sequence permutation mechanism, we obtain that $%
R_{1}=\{\lambda ^{\left( a\right) },\lambda ^{\left( b\right) },\lambda
^{\left( c\right) }\}$, $R_{2}=\{\lambda ^{\left( b\right) },\lambda
^{\left( c\right) },\lambda ^{\left( a\right) }\}$ and $R_{3}=\{\lambda
^{\left( c\right) },\lambda ^{\left( a\right) },\lambda ^{\left( b\right)
}\} $. Based on the mode status matrix, for the selection of $R_{1}$, we
obtain $\left\vert 0\right) \otimes \left\vert 0\right) \otimes \left\vert
0\right) $; for the selection of $R_{2}$, we obtain $\left\vert 1\right)
\otimes \left\vert 1\right) \otimes \left\vert 1\right) $; for the selection
of $R_{3}$, we obtain nothing. Thus we could reconstruct the state based on
the mode status matrix%
\begin{eqnarray}
\left\vert \Psi _{GHZ}\right) &=&\frac{1}{\sqrt{2}}\sum_{r=1}^{3}\left[
\left( \tilde{\alpha}_{a}^{j\left[ R_{r}^{a}\right] }\left\vert 0\right) +%
\tilde{\beta}_{a}^{j\left[ R_{r}^{a}\right] }\left\vert 1\right) \right)
\otimes \left( \tilde{\alpha}_{b}^{j\left[ R_{r}^{b}\right] }\left\vert
0\right) +\tilde{\beta}_{b}^{j\left[ R_{r}^{b}\right] }\left\vert 1\right)
\right) \otimes \left( \tilde{\alpha}_{c}^{j\left[ R_{r}^{c}\right]
}\left\vert 0\right) +\tilde{\beta}_{c}^{j\left[ R_{r}^{c}\right]
}\left\vert 1\right) \right) \right]  \nonumber \\
&=&\frac{1}{\sqrt{2}}\left( \left\vert 000\right) +\left\vert 111\right)
\right) .  \label{eq34}
\end{eqnarray}

\subsubsection{Classical simulation of W state \label{sec3.3.4}}

Then we discuss the classical simulation of W state,%
\begin{equation}
\left\vert \Psi _{W}\right\rangle =\frac{1}{\sqrt{3}}\left( \left\vert
1_{a}\right\rangle \left\vert 0_{b}\right\rangle \left\vert
0_{c}\right\rangle +\left\vert 0_{a}\right\rangle \left\vert
1_{b}\right\rangle \left\vert 0_{c}\right\rangle +\left\vert
0_{a}\right\rangle \left\vert 0_{b}\right\rangle \left\vert
1_{c}\right\rangle \right) ,  \label{eq35}
\end{equation}%
which contains three classical fields as follows
\begin{eqnarray}
\left\vert \psi _{a}\right) &=&\frac{1}{\sqrt{3}}\left( e^{i\lambda ^{\left(
a\right) }}\left\vert 1\right) +e^{i\lambda ^{\left( b\right) }}\left\vert
0\right) +e^{i\lambda ^{\left( c\right) }}\left\vert 0\right) \right) ,
\label{eq36} \\
\left\vert \psi _{b}\right) &=&\frac{1}{\sqrt{3}}\left( e^{i\lambda ^{\left(
a\right) }}\left\vert 1\right) +e^{i\lambda ^{\left( b\right) }}\left\vert
0\right) +e^{i\lambda ^{\left( c\right) }}\left\vert 0\right) \right) ,
\nonumber \\
\left\vert \psi _{c}\right) &=&\frac{1}{\sqrt{3}}\left( e^{i\lambda ^{\left(
a\right) }}\left\vert 1\right) +e^{i\lambda ^{\left( b\right) }}\left\vert
0\right) +e^{i\lambda ^{\left( c\right) }}\left\vert 0\right) \right) ,
\nonumber
\end{eqnarray}%
It is noteworthy that the three classical fields could be produced from one
single field by using two beam splitters, which is quite similar with the
generation of W state in quantum mechanics. Performing the same scheme, we
obtain the mode status matrix%
\begin{equation}
M\left( \tilde{\alpha}_{i}^{j},\tilde{\beta}_{i}^{j}\right) =\left(
\begin{array}{ccc}
\left( 0,1\right) & \left( 1,0\right) & \left( 1,0\right) \\
\left( 0,1\right) & \left( 1,0\right) & \left( 1,0\right) \\
\left( 0,1\right) & \left( 1,0\right) & \left( 1,0\right)%
\end{array}%
\right) .  \label{eq37}
\end{equation}%
According to the sequence permutation mechanism, we obtain that $%
R_{1}=\{\lambda ^{\left( a\right) },\lambda ^{\left( b\right) },\lambda
^{\left( c\right) }\}$, $R_{2}=\{\lambda ^{\left( b\right) },\lambda
^{\left( c\right) },\lambda ^{\left( a\right) }\}$ and $R_{3}=\{\lambda
^{\left( c\right) },\lambda ^{\left( a\right) },\lambda ^{\left( b\right)
}\} $ again. Based on the mode status matrix, we obtain $\left\vert 1\right)
\otimes \left\vert 0\right) \otimes \left\vert 0\right) $, $\left\vert
0\right) \otimes \left\vert 0\right) \otimes \left\vert 1\right) $, $%
\left\vert 0\right) \otimes \left\vert 1\right) \otimes \left\vert 0\right) $
for the selection of $R_{1}$, $R_{2}$, $R_{3}$, respectively. We find an
interesting fact that when we obtain the state $\left\vert 1\right) $ of the
first field, $R_{1}$ must be selected, thus only the $\left\vert 0\right)
\otimes \left\vert 0\right) $ state could be obtained from the other two
fields; otherwise when we obtain the state $\left\vert 0\right) $ of the
first field, the selection could be $R_{2}$ or $R_{3}$, thus the state of $%
\left\vert 0\right) \otimes \left\vert 1\right) +\left\vert 1\right) \otimes
\left\vert 0\right) $ could be obtained from the other two fields. This fact
is quite similar with the case of quantum measurement and the collapse
phenomenon for W state in quantum mechanics. We could reconstruct the state
based on the mode status matrix%
\begin{eqnarray}
\left\vert \Psi _{W}\right) &=&\frac{1}{\sqrt{3}}\sum_{r=1}^{3}\left[ \left(
\tilde{\alpha}_{a}^{j\left[ R_{r}^{a}\right] }\left\vert 0\right) +\tilde{%
\beta}_{a}^{j\left[ R_{r}^{a}\right] }\left\vert 1\right) \right) \otimes
\left( \tilde{\alpha}_{b}^{j\left[ R_{r}^{b}\right] }\left\vert 0\right) +%
\tilde{\beta}_{b}^{j\left[ R_{r}^{b}\right] }\left\vert 1\right) \right)
\otimes \left( \tilde{\alpha}_{c}^{j\left[ R_{r}^{c}\right] }\left\vert
0\right) +\tilde{\beta}_{c}^{j\left[ R_{r}^{c}\right] }\left\vert 1\right)
\right) \right]  \nonumber \\
&=&\frac{1}{\sqrt{3}}\left( \left\vert 100\right) +\left\vert 010\right)
+\left\vert 001\right) \right) .  \label{eq38}
\end{eqnarray}

\section{Generalization of the simulation and some discussions on efficiency
\label{sec4}}

In this section, we first propose a generalization of our simulation to the
case of an arbitrary quantum state, then the efficiency of our simulation is
discussed. Besides, for better understanding our simulation, the PPSs in the
cases of modulating two and three classical fields is illustrated.

\subsection{Generalization of the simulation \label{sec4.1}}

Here we propose a generalization of our simulation to the case of an
arbitrary quantum state. In Sec. \ref{sec3.3}, classical simulation of
several typical quantum states has been introduced, including product state,
Bell states, GHZ state, and W state, based on which we assume that any
quantum state of $n-1$ particles could be successfully simulated. Here we
consider the classical simulation of an arbitrary quantum state of $n$
particles, $\left\vert \Psi _{n}\right\rangle $. We first transform $%
\left\vert \Psi _{n}\right\rangle $ to an equivalent expression, $\left\vert
\Psi _{n}\right\rangle =\left\vert \Phi _{n-1}\right\rangle \left\vert
0\right\rangle _{n}+\left\vert \Theta _{n-1}\right\rangle \left\vert
1\right\rangle _{n}$, where $\left\vert \Phi _{n-1}\right\rangle $ and $%
\left\vert \Theta _{n-1}\right\rangle $ denote two quantum states of $n-1$
particles. According to the assumption, we obtain that the quantum state $%
\left\vert \Phi _{n-1}\right\rangle $ could be simulated by $n-1$ classical
fields $\left\vert \varphi _{1}\right) ,\ldots ,\left\vert \varphi
_{n-1}\right) $ modulated with $n-1$ PPSs $\left\{ \lambda ^{\left( 1\right)
},\ldots ,\lambda ^{\left( n-1\right) }\right\} $, and the quantum state $%
\left\vert \Theta _{n-1}\right\rangle $ could be simulated by $n-1$
classical fields $\left\vert \vartheta _{1}\right) ,\ldots ,\left\vert
\vartheta _{n-1}\right) $ modulated with $n-1$ PPSs $\left\{ \lambda
^{\left( 1\right) },\ldots ,\lambda ^{\left( n-2\right) },\lambda ^{\left(
n\right) }\right\} $.

Further, we consider the classical simulation of the quantum state $%
\left\vert \Psi _{n}\right\rangle $, which is the superposition of the
classical simulation of $\left\vert \Phi _{n-1}\right\rangle \otimes
\left\vert 0\right\rangle _{n}$ and the classical simulation of $\left\vert
\Theta _{n-1}\right\rangle \otimes \left\vert 1\right\rangle _{n}$,%
\begin{eqnarray}
\left\vert \psi _{1}\right) &=&\frac{1}{C_{1}}[\left\vert \varphi
_{1}\right) +\left\vert \vartheta _{1}\right) ],  \label{eq41} \\
&&......  \nonumber \\
\left\vert \psi _{n-1}\right) &=&\frac{1}{C_{n-1}}[\left\vert \varphi
_{n-1}\right) +\left\vert \vartheta _{n-1}\right) ],  \nonumber \\
\left\vert \psi _{n}\right) &=&\frac{1}{C_{n}}\left[ e^{i\lambda ^{\left(
n\right) }}\left\vert 0\right) +e^{i\lambda ^{\left( n-1\right) }}\left\vert
1\right) \right] .  \nonumber
\end{eqnarray}

Thus by only adding one classical field and one PPS, we obtain the classical
simulation of the quantum state $\left\vert \Psi _{n}\right\rangle $ based
on the assumed classical simulation of $\left\vert \Phi _{n-1}\right\rangle $
and $\left\vert \Theta _{n-1}\right\rangle $. Using the principle of
induction, we could provide classical simulation of any quantum state.
Therefore we successfully generalize our simulation.

\subsection{Efficiency of the classical simulation \label{sec4.2}}

For a long time, researchers have used classical fields to simulate quantum
states and quantum computation. In these researches, multiple qubits are
distinguished by different degrees of freedom, such as optical modes or
space positions. However, as no tensor product structure is obtained in
these classical simulations, each quantum base state needs independent
degree of freedom to simulate. In quantum mechanics, the number of quantum
base states grows exponentially with the number of quantum particles.
Therefore the classical simulation require resources that also grow
exponentially with the number of simulated quantum particles, which is not
efficient.

In this paper, we utilize the properties of PPSs to distinguish classical
fields that are even overlapped in same space and time. A $2^{n}$%
-dimensional Hilbert space which contains tensor product structure is
spanned by $n$\ classical fields modulated with PPSs. In our scheme, the
resources required are classical fields modulated with PPSs instead of
optical/space modes. One classical field modulated with one PPS can simulate
one quantum particle. It means that the amount of classical fields and PPSs
grows linearly with the number of quantum particles. According to the
m-sequence theory, the number of PPSs in the set $\Xi $ equals to the length
of sequences, which means that the time resource (the length of sequence)
required also grows linearly with the number of the particles. Based on the
analysis above, we conclude that one can efficiently simulate quantum
entanglement with linearly growing resources by using our scheme.

\subsection{Illustration of some pseudorandom phase sequences \label{sec4.3}}

For better understanding our scheme, the PPSs in the cases of modulating two
and three classical fields is illustrated below. Using the method mentioned
in section \ref{sec2}, an m-sequence of length $4^{2}-1$\ is generated by a
primitive polynomial of the lowest degree over $GF(4)$, which is $\left[
\begin{array}{ccccccccccccccc}
1 & 2 & 0 & 3 & 3 & 2 & 3 & 0 & 1 & 1 & 3 & 1 & 0 & 2 & 2%
\end{array}%
\right] $. Further we obtain a group that includes 16 PPSs of length 16: $%
\left\{ \lambda ^{\left( 0\right) },\ldots ,\lambda ^{\left( 15\right)
}\right\} $, of which in exception to $\lambda ^{\left( 0\right) }$, all
PPSs are independent and could be used to modulate classical fields to
simulate quantum states of up to 15 particles. We could choose any two PPSs
from the group for the simulation of two particles quantum state, for
example,%
\begin{eqnarray}
\lambda ^{\left( a\right) } &=&\left[
\begin{array}{cccccccccccccccc}
1 & 2 & 0 & 3 & 3 & 2 & 3 & 0 & 1 & 1 & 3 & 1 & 0 & 2 & 2 & 0%
\end{array}%
\right] \times \pi /2,  \label{eq70} \\
\lambda ^{\left( b\right) } &=&\left[
\begin{array}{cccccccccccccccc}
2 & 1 & 2 & 0 & 3 & 3 & 2 & 3 & 0 & 1 & 1 & 3 & 1 & 0 & 2 & 0%
\end{array}%
\right] \times \pi /2.  \nonumber
\end{eqnarray}%
And we could also choose any three PPSs from the group for simulation of
three particles quantum state, for example,%
\begin{eqnarray}
\lambda ^{\left( a\right) } &=&\left[
\begin{array}{cccccccccccccccc}
1 & 2 & 0 & 3 & 3 & 2 & 3 & 0 & 1 & 1 & 3 & 1 & 0 & 2 & 2 & 0%
\end{array}%
\right] \times \pi /2,  \label{eq71} \\
\lambda ^{\left( b\right) } &=&\left[
\begin{array}{cccccccccccccccc}
2 & 1 & 2 & 0 & 3 & 3 & 2 & 3 & 0 & 1 & 1 & 3 & 1 & 0 & 2 & 0%
\end{array}%
\right] \times \pi /2,  \nonumber \\
\lambda ^{\left( c\right) } &=&\left[
\begin{array}{cccccccccccccccc}
2 & 2 & 1 & 2 & 0 & 3 & 3 & 2 & 3 & 0 & 1 & 1 & 3 & 1 & 0 & 0%
\end{array}%
\right] \times \pi /2.  \nonumber
\end{eqnarray}%
Using classical fields modulated the PPSs given above, we can efficiently
simulate any quantum state of two or three particles.

\section{Conclusions \label{sec5}}

In this paper, we have discussed a new scheme to simulate quantum states by
using classical fields modulated with pseudorandom phase sequences. We first
demonstrated that $n$ classical fields modulated with $n$\ different PPSs
can constitute a $2^{n}$-dimensional Hilbert space that contains tensor
product structure, which is similar with quantum systems. Further, by
performing quadrature demodulation scheme, we obtained the mode status
matrix of the simulating classical fields, based on which we proposed a
sequence permutation mechanism to reconstruct the simulated quantum states%
\textbf{.} Besides, classical simulation of several typical quantum states
was discussed, including product state, Bell states, GHZ state and W state.
We generalized our simulation and discussed the efficiency of our simulation
finally. We conclude that quantum states can be efficiently simulated by
using classical fields modulated with pseudorandom phase sequences. The
research on simulation of quantum states may be important, for it not only
provides useful insights into fundamental features of quantum mechanics, but
also yields new insights into quantum computation and quantum communication.

Supported by the National Natural Science Foundation of China under Grant No
60407003, and National Basic Research Program 973 of China under Grant
2007CB307003. The authors thank Professors LIU Xu and TONG Li-Min for
helpful discussion.

\begin{description}
\item \newpage

\item[Fig. 1] The PPS encoding scheme for one input field, where PNG denotes
the pseudorandom number generator and PM denotes the phase modulator.

\item[Fig. 2] The PPS quadrature demodulation scheme for one input field.

\item[Fig. 3] The PPS quadrature demodulation scheme for one field with two
orthogonal modes, where the gray block denote the mode splitter.

\item[Fig. 4] The PPS quadrature demodulation scheme for multiple input
fields, where the M block is shown in Fig. 3.

\item[Fig. 5] The mode status matrix related to the fields and PPSs, where $%
M_{i}^{j}$ is the element of $M\left( \tilde{\alpha}_{i}^{j},\tilde{\beta}%
_{i}^{j}\right) $ for the $i$th classical field and the reference PPS $%
\lambda ^{\left( j\right) }$.
\end{description}

\end{document}